\def\comment#1{}

\documentclass[aps,prl,showpacs,reprint]{revtex4-1}
\usepackage{graphicx}

\begin{document}


\title{Electrical suppression of spin relaxation in GaAs(111)B Quantum Wells} 



\author{A. Hern\'andez-M\'inguez}
\email[e-mail address: ]{alberto@pdi-berlin.de}
\author{K. Biermann}
\author{R. Hey}
\author{P. V. Santos}
\affiliation{Paul-Drude-Institut f\"ur Festk\"orkperelektronik, Hausvogteiplatz 5-7, 10117 Berlin, Germany}



\begin{abstract}
Spin dephasing via the spin-orbit interaction (SOI) is a major mechanism limiting the electron spin lifetime in III-V zincblende quantum wells. The dephasing can be suppressed in GaAs(111) quantum wells by applying an electric field. The suppression has been attributed to the compensation of the intrinsic SOI associated by the bulk inversion asymmetry (BIA) of the GaAs lattice by a structural induced asymmetry (SIA) SOI term induced by an electric field. We provide direct experimental evidence for this mechanism by demonstrating the transition between the BIA-dominated to a SIA-dominated regime via photoluminescence measurements carried out over a wide range of applied fields. Spin lifetimes exceeding 100~ns are obtained near the compensating electric field, thus making GaAs (111) QWs  excellent candidates  for the  electrical storage and manipulation of spins.

\end{abstract}

\pacs{72.25.Rb, 72.25.Fe, 78.67.De}

\maketitle 

The manipulation of electron spins in semiconductors has attracted much interest in recent years due to potential applications for quantum  information processing~\cite{Awschalom2002,Zutic_ROMP76_323_04}. One obstacle towards these applications are the short spin lifetimes, which are presently behind the values required for efficient spin storage and manipulation. Improvement of the electron spin lifetime in III-V semiconductor quantum wells (QWs) demands the control of relaxation mechanisms associated with the spin-orbit interaction (SOI). The impact of this interaction on spins can be described in terms of an effective magnetic field, whose  magnitude and orientation depend on the electronic wave vector, $\mathbf{k_\parallel}=(k_x, k_y)$. Spins from an initially aligned ensemble  moving with different $\mathbf{k}$'s will precess with different Larmor frequencies, $\mathbf{\Omega}_\mathrm{SO}$, leading to a loss of the initial spin polarization within a few nanoseconds -- a phenomenon known as the  Dyakonov-Perel (DP) spin dephasing mechanism~\cite{Dyakonov_SPSS13_3023_72,Dyakonov_SPS20_110_86}.

In high quality GaAs QWs,  the $\mathbf{\Omega}_\mathrm{SO}$ is dominated by two major contributions. The first arises from the intrinsic bulk inversion asymmetry (BIA, also know as the Dresselhaus~\cite{Dresselhaus55a} term) of the III-V lattice. The second, denoted as the structure induced asymmetry (SIA, or Rashba~\cite{Bychkov1984}) term, appears as a consequence of a symmetry reduction imposed by an external perturbation. The latter can be controlled by an electric field applied along the  QW growth direction, $z$.  The impact of these contributions on the electron spin dynamics depends on dimensionality as well as on the QW symmetry, which is defined by its crystallographic growth direction~\cite{Dyakonov_SPS20_110_86}. Of special interest for long spin lifetimes are (111) QWs, where the BIA ($\mathbf{\Omega}_\mathrm{BIA}$) and SIA ($\mathbf{\Omega}_\mathrm{SIA}$) precession frequencies  can be expressed as~\cite{Winkler03a,Cartoixa05a}:

\begin{eqnarray}
\label{Eq1}
\mathbf{\Omega}_\mathrm{BIA}(\mathbf{k}_{\parallel})&=&\frac{2\gamma}{\hbar\sqrt{3}}
	\left[
	 \left( \langle k_z^2\rangle - \frac{1}{4}  k_{\parallel}^2  \right)   (k_y,-k_x,0) \right. + \\ \nonumber
  & &\left.  (0,0,\frac{\sqrt{2}}{4} k_y  (k_y^2-3k_x^2))  	\right] \\
\label{Eq2}
\mathbf{\Omega}_\mathrm{SIA}(\mathbf{k}_{\parallel})&=&\frac{2E_zr_{41}}{\hbar}(k_y,-k_x,0),
\end{eqnarray}

\noindent respectively, where the reference frame is defined by the axes $x=\frac{1}{\sqrt{6}}[\bar1\bar12]$, $y=\frac{1}{\sqrt{2}}[1\bar10]$, and $z=\frac{1}{\sqrt{3}}[111]$. Here, $\langle k^2_z \rangle = \left( \pi/d_\mathrm{eff}\right)^2$ is the averaged squared wave vector along $z$, which is determined by the spatial extension $d_\mathrm{eff}$ of the electronic wave function.  $\gamma$ and  $r_{41}$ are the Dresselhaus and Rashba coefficient, respectively,  and $E_z$ is the amplitude of the electric field applied across the structure.
At low temperature, quadratric terms in $k_\parallel$ can be neglected. In this case, both the BIA and the SIA terms become parallel to each other and lie in the QW plane. Moreover, by choosing

\begin{equation}\label{comp_point}
E_z=E_c=\frac{\gamma}{\sqrt{3}r_{41}} \left( \langle k_z^2\rangle - \frac{1}{4}  k_{\parallel}^2  \right)
\label{EqEc}
\end{equation}

\noindent in order to make $\mathbf{\Omega}_\mathrm{SIA}=-\mathbf{\Omega_\mathrm{BIA}}$, it becomes possible to suppress DP relaxation by linear terms in k for \textit{all} spin orientations~\cite{Cartoixa05a}, leading to very long spin lifetimes.

\begin{figure}
\includegraphics[width=\columnwidth]{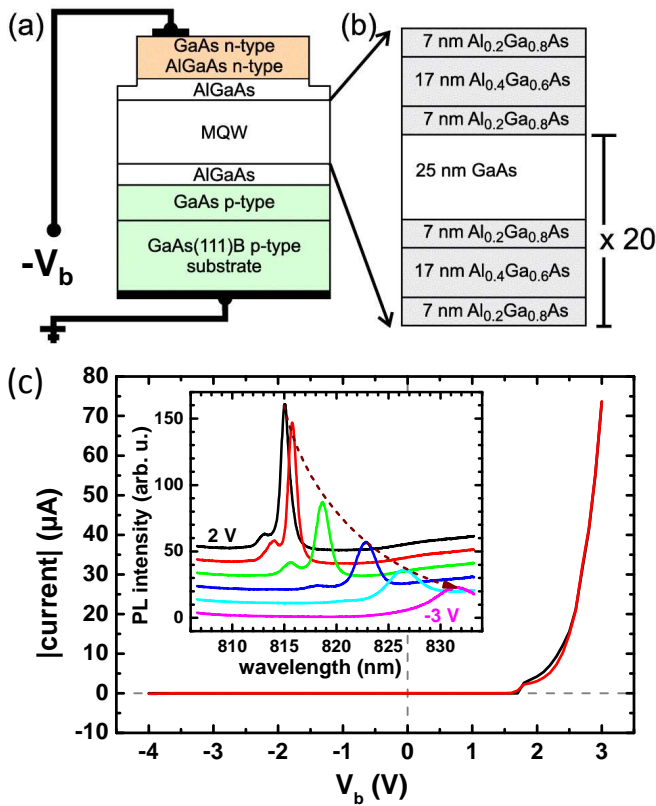}
\caption{(color online) (a) Structure of a multiple quantum well (MQW) embedded in a n-i-p sample and the electric contacts. The bias voltage, $V_b$,  is applied between the top  $n$ region and the $p$-doped substrate. (b) Layer structure of the active region of the MQW sample. (c) $I$ {\it vs.} $V_b$ curves of the single quantum well (SQW) at $T=39$~K, without (black) and with (red) illumination. Inset: photoluminescence (PL) of the SQW for $V_b$ going from 2~V (black curve) to -3~V (pink curve) in steps of -1~V. \label{structure}}
\end{figure}

Recently, Balocchi {\it et al.}~\cite{Balocchi_PRL107_136604_11,Ye_APL101_32104_12} and Biermann {\it et al.}~\cite{PVS257} have reported on the enhancement of the spin lifetime in  GaAs (111) QWs subjected to an external field, which was attributed to the previously mentioned SO compensation. The applied electric fields, however, were below the  $E_c$ values expected from Eq.~\ref{comp_point}. The  transition from a BIA-dominated to a SIA-dominated, which would unambiguously demonstrate the compensation mechanism and, simultaneously, provide the longest spin lifetimes, has, to our knowledge, so far not been experimentally established. A main obstacle to reach the transition is associated with the high values for $E_c$, which can induce carrier extraction from the QW by field ionization. In addition, a strong field reduces, via the quantum confined Stark effect (QCSE), the spatial overlap of the electron and hole wave functions. While the latter reduces spin relaxation via the electron-hole exchange interaction~\cite{Bir_SPJ42_705_76}, it also considerably  diminishes the radiative recombination rate and, consequently, the yield of the polarization-resolved photoluminescence (PL) technique normally used for probing spins.

In this contribution, we overcome this limitation by probing spins via a novel PL technique combining pulsed illumination and biasing. This technique enables measurements of the spin lifetime in (111) GaAs QWs embedded in n-i-p diode structures over a wide range of electric fields across the compensation point. We show that the lifetime $\tau_z$ of optically injected, z-oriented spins initially increases with reverse bias, reaches a maximum, and then reduces for higher biases. The maximum is attributed to the transition between a regime dominated by the BIA contribution to another determined by the bias-induced SIA term: its observation provides a conclusive evidence of the SOI compensation mechanism. The measured compensation field $E_c$ yields ratios $r_{14}/\gamma$ (cf.~Eq.~\ref{EqEc}) comparable to the ones reported in the literature~\cite{Balocchi_PRL107_136604_11,Walser2012}. Moreover, the spin lifetime close to $E_c$ reaches values exceeding 100~ns, which are among the highest reported for GaAs structures. The bias also enhances the lifetime $\tau_R$ of spins precessing around an in-plane magnetic field $B$, but not to the same levels as $\tau_z$. Studies of the temperature dependence prove that the shorter lifetimes $\tau_R$ near $E_c$ cannot be accounted by the higher order terms in k of  $\mathbf{\Omega}_\mathrm{BIA}(\mathbf{k}_{\parallel})$ in Eq.~1, thus indicating the existence of an additional relaxation channel for in-plane spins.

The studies were carried out on two kinds of samples. The first consists of a GaAs multiple quantum well (MQW) grown by molecular beam epitaxy on a p-doped GaAs(111)B substrate tilted by $\delta\theta=2^\circ$  towards $x$. As $\langle k^2_z \rangle$ in Eq.~\ref{comp_point} is proportional to $d_\mathrm{eff}^{-2}$, the compensation field reduces for thick QWs. We took advantage of this fact by using thick (25 nm-thick) GaAs QWs. The MQW sample [cf.~Fig.~\ref{structure}(a)] consists of a stack of 20 QWs separated by (Al,Ga)As barriers [c.f. Fig.~\ref{structure}(b)]. In order to apply the electric control field, the MQW stack  was embedded within the intrinsic region of a n-i-p structure.
The bias voltage ($V_{b}$) was applied between an Al Schottky contact deposited on top of the structure and the p-type doped substrate. To confine the applied voltage along the z-direction, the top doped layers and part of the top (undoped) (Al,Ga)As spacer layer above the MQW were processed into mesa structures with a diameter of 300~$\mu$m by wet chemical etching.  The second sample contains a single QW (SQW) grown on a substrate with $\delta\theta=1^\circ$. Its structure is similar to the MQW in Fig.~\ref{structure}(a) with the difference that all QWs, except the one in the middle of the stack, were replaced by (Al,Ga)As barriers.

The time-resolved PL studies were performed at different temperatures in a cold finger cryostat with a  window for optical access and electric feed-throughs for the application of high-frequency bias pulses. Care was taken during sample mounting to reduce the build-up of stress fields in the sample during the cooling process. Spin polarized charge carriers were selectively excited in the QWs using a circularly polarized pulsed laser beam with a wavelength of 757~nm and a repetition period of 80~ns focused on the sample surface by a microscope objective. The PL emitted by the QWs was collected by the same objective, spectrally filtered by band-pass filters, and split into two beams with intensity proportional to its left ($I_L$) and right ($I_R$) circular components using a quarter-wave plate followed by a polarizing beam splitter. The two beams were then detected with time resolution by a pair of photodetectors synchronized with the laser pulses. From the detected signals, we determined the temporal evolution of the electron spin polarization defined as $\rho_s(t)=[I_R(t)-I_L(t)]/[I_R(t)+I_L(t)]$. Although the relaxation process from the optically excited high energy states to their quasi thermal equilibrium ones implies some reduction of the initial orientation of the electron spins, a significant fraction of the optically generated electron spin polarization is preserved in wide QWs~\cite{Pfalz_PRB71_165305_05}. Therefore, as the typical hole spin lifetimes in the present temperature range are limited to a few hundreds of picoseconds~\cite{Damen1991}, the polarization of the photons emitted for longer times via electron-hole recombination is determined only by the electron spin orientation.

\begin{figure}
\includegraphics[width=\columnwidth]{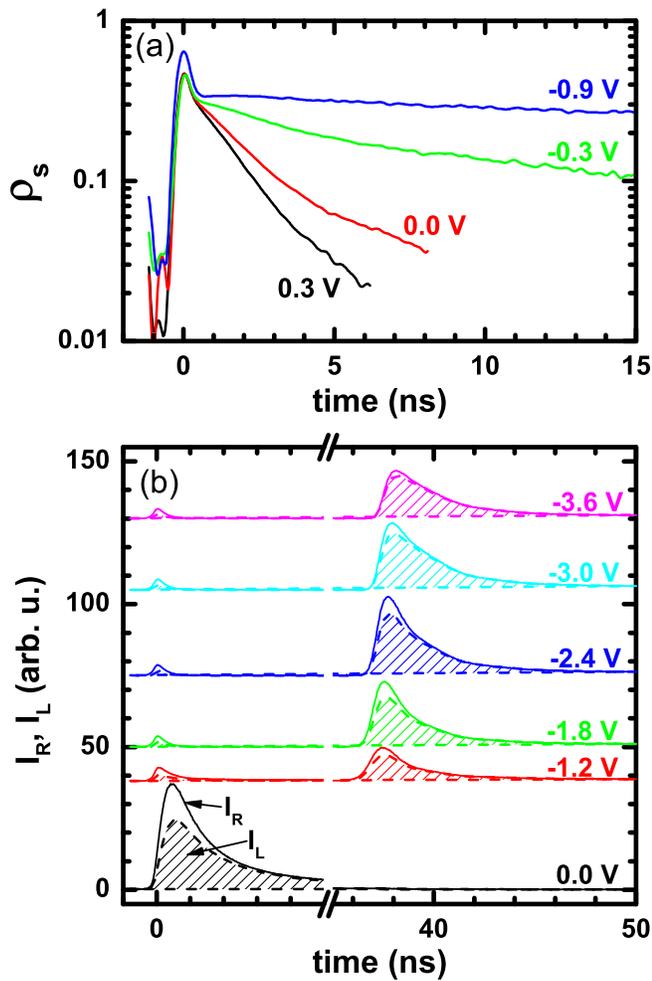}
\caption{(color online) (a) Time-resolved spin polarization, $\rho_s$, of the MQW for different bias voltages $V_b>-1.2$~V. (b) Right ($I_R$, solid lines) and  left ($I_L$, dashed lines) circularly polarized PL as a function of time for voltage pulses of 40~ns, repetition period of 80~ns and amplitude $V_b\leq-1.2$~V. The lowest curve shows the time dependence of the  $I_R$ and $I_L$ recorded for $V_b=0$.~\label{spin1}}
\end{figure}

Figure~\ref{structure}c shows the $I$ {\it vs.} $V_b$ curves of the SQW recorded at 39~K  in the dark (black curve) and under illumination (red curve). Note that according to the convention of Fig.~\ref{structure}(a), forward (reverse) bias corresponds to $V_{b}>0$ ($V_{b}<0$). The spectroscopic experiments were carried out for biases between -3.5 and 2~V, where the current through the diodes is negligible. The PL spectra of the SQW displayed in the inset show that the energy, line width, and intensity of the PL line change due to the QCSE induced by the applied bias. Figure~\ref{spin1}(a) shows the spin polarization dynamics of the MQW in the region of low reverse bias (for $V_b>-1.2$~V). The spin lifetime increases considerably with reverse bias, in accordance with results already reported \cite{Balocchi_PRL107_136604_11,Ye_APL101_32104_12,PVS257}. This behavior is attributed to the partial compensation of the BIA term by the SIA term induced by the externally applied electric field. Note, however, that the PL intensity reduces substantially with bias due to the QCSE. This hinders the detection of spins at the large reverse fields required for compensation.

The previous limitation was overcome by carrying out the experiments under pulsed reverse bias. Here, laser and bias pulses are synchronized and have the same repetition rate. The laser pulse hits the sample at the time instant $t=0$ shortly after the application of a reverse bias pulse of variable amplitude $V_b$. As in the previous experiments, the reverse bias  prevents the recombination of the photoexcited spin polarized carriers, which remain stored in the QWs during the bias pulse. Information about the density of stored carriers and spins is extracted at the end of the bias pulse, when the diode structure is subjected to a small forward bias ($V_b=1$~V) to induce carrier recombination. $I_R(t)$ and $I_L(t)$  traces recorded under a 40~ns long bias pulse with a repetition period of 80~ns are illustrated in Fig.~\ref{spin1}(b). The amplitude of the PL after the bias pulse increases for $V_b<-1.2$~V as the carrier lifetime exceeds the pulse width. The PL  rise time is determined by the falling time of the bias pulses of $t_r=2$~ns. For pulse biases $V_b<-2$~V, the integrated PL after the pulse corresponds to almost 60\% of the one detected  right after the laser pulse under a dc-bias of 0~V [lowest curve in Fig.~\ref{spin1}(b)], thus indicating that the QWs can efficiently store a high density of carriers over long times. The reduction of the retrieved PL intensities for pulse voltages $V_b<-3.5$~V is attributed to field-induced carrier extraction.

\begin{figure*}
\includegraphics[width=\textwidth]{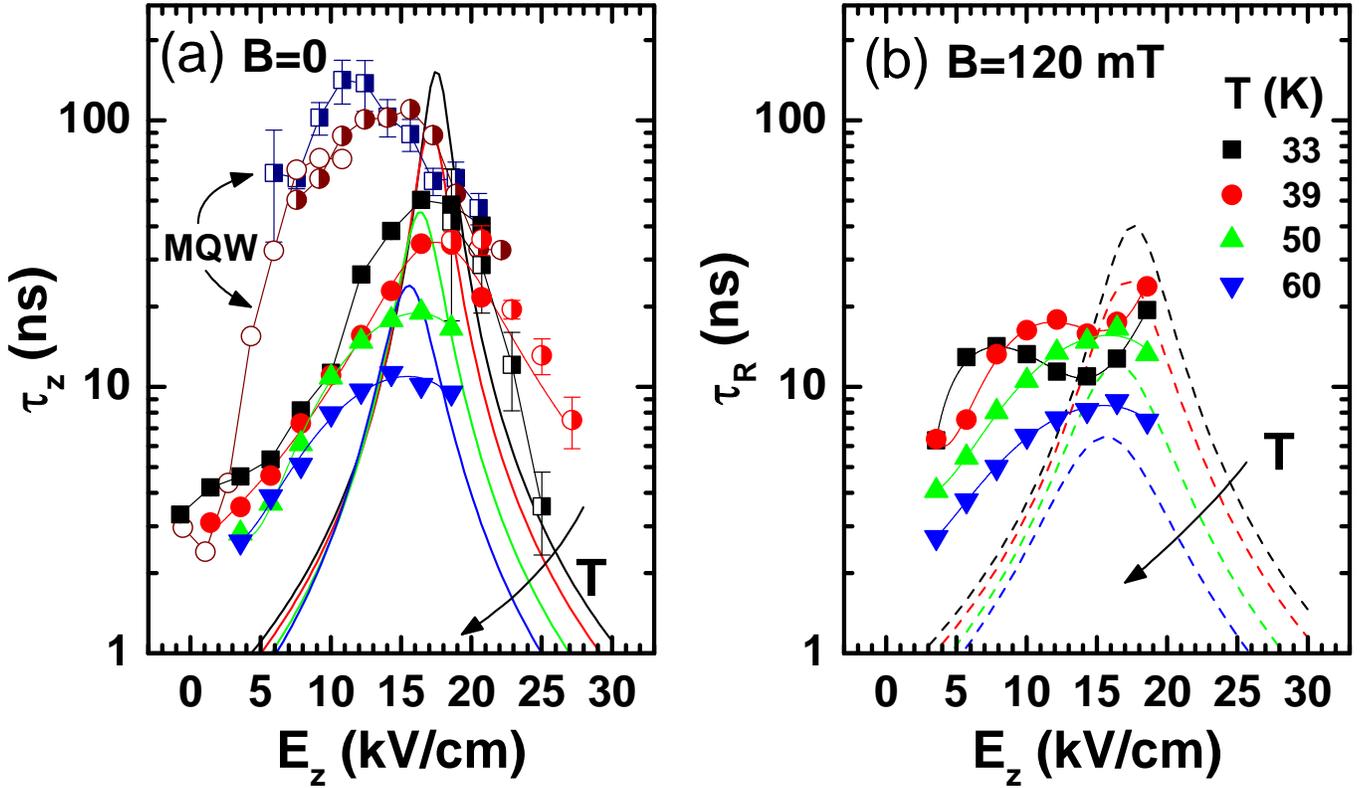}
\caption{(color online) Spin lifetime as a function of the applied electric field ($E_z$) and temperature  (a) in the absence ($\tau_z$), and (b) in the presence of an in-plane magnetic field $B=120$~mT ($\tau_R$). The results were obtained from the decay of the spin polarization just after the laser pulse (filled and open symbols) and after the bias pulse (half filled symbols). The curves marked MQW in (a) are for the MQW at 33~K (dark blue squares) and 39~K (dark red circles). The measurement temperatures for the SQW are:  33~K (black squares), 39~K (red circles), 50~K (green up triangles), 60~K (blue down triangles). The solid and dashed lines are the lifetimes calculated at the same temperatures using the model described in detail in the text for $B=0$ and $B=120$~mT, respectively.\label{spinlifetime}}
\end{figure*}

The remarkable difference in intensity of the $I_R$ and $I_L$ light pulses in Fig.~\ref{spin1}(b)  after a delay of almost 40~ns attests to the conservation of the spin polarization during charge storage. Note that this difference vanishes shortly after the bias pulse due to the reduced spin lifetime under forward bias. The lifetime of the stored spins was determined by recording $I_R(t)$ and $I_L(t)$ traces for different bias pulse widths and calculating the decay rate of spin polarization defined as before. The half filled symbols in Fig.~\ref{spinlifetime}(a) summarize $\tau_z$ measured in the MQW and SQW samples under pulsed biases at different temperatures. The data matches well the corresponding results for low electric fields obtained from the decay of the spin polarization directly after the laser pulse (solid and open symbols). In all cases, the field dependence of the spin lifetime shows a maximum for fields of $15.0\pm0.3$ and $18.5\pm0.3$~kV/cm for the MWQ and SQW, respectively. The observation of these maxima unambiguously establishes the BIA/SIA compensation as the mechanism for spin enhancement in (111) QWs. Moreover, $\tau_z$  increases with decreasing temperature and  exceeds 50 and  100~ns close to $E_c$ for the SQW and MQW samples: these lifetimes are among the longest values reported for GaAs structures. The extension of the SIA-dominated bias region beyond the compensation point is limited up to values of the electric field of about $1.5-2$ times longer than the required one for compensation. This maximum applied field is determined by field-induced carrier extraction under high reverse biases.

\begin{figure}
\includegraphics[width=\columnwidth]{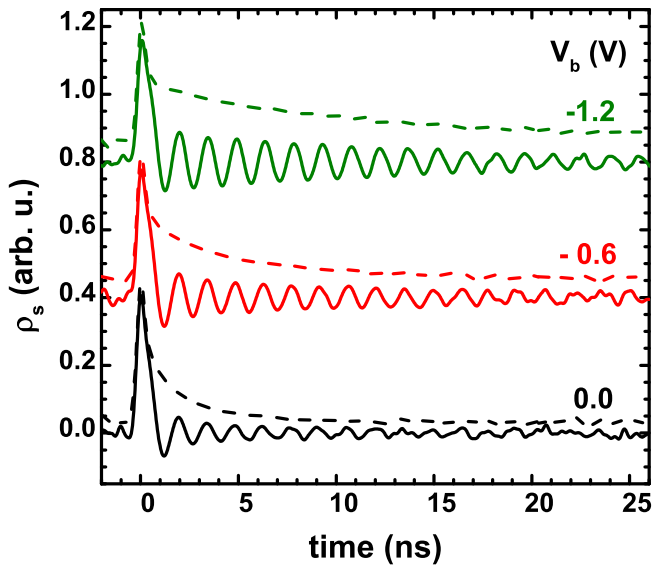}
\caption{(color online) (a) Spin dynamics in the SQW for three different bias voltages $V_b$ measured at 39~K under $B=120$~mT (solid lines) and in the absence of external magnetic field (dashed lines).\label{spin2}}
\end{figure}

According to Eq.~\ref{EqEc}, the compensation field $E_c$ depends on the ratio $r_{41}/\gamma$ between Rashba and Dresselhaus spin-splitting constants. The effective QW thickness of $d_\mathrm{eff}=28.6$~nm was determined by  calculating the penetration  of the electronic wave function into the (Al,Ga)As barriers using a {\bf k.p} model. For these wide QWs, the average thermal values $\langle k_\parallel^2\rangle \sim 0.25 \langle k_z^2\rangle$ at $T>30$~K cannot be neglected and the higher order terms in Eq.~\ref{Eq1} have to be taken into account. For this purpose, we have calculated the DP scattering rate ${\tau_i}$ for spins along $i$ according to~\cite{Cartoixa05a}:

\begin{equation}
\frac{1}{\tau_i} = {\tau^*_p} \left( \langle\Omega^2_{\mathrm{SO},j} (k)\rangle + \langle\Omega^2_{\mathrm{SO},k} (k)\rangle \right), \qquad i\neq j\neq k=x, y, z.
\label{Eqtaui}
\end{equation}

\noindent Here, $\tau^*_p$ is proportional to the momentum relaxation time and $\mathbf{\Omega_\mathrm{SO}}(\mathbf{k}_{\parallel})=\mathbf{\Omega}_\mathrm{BIA}(\mathbf{k}_{\parallel}) + \mathbf{\Omega}_\mathrm{SIA}(\mathbf{k}_{\parallel}) + \mathbf{\Omega}^{(\theta)}_\mathrm{BIA}(\mathbf{k}_{\parallel})$, where we include in addition to the terms in Eqs.~\ref{Eq1} and \ref{Eq2} the correction $\mathbf{\Omega}^{(\theta)}_\mathrm{BIA}(\mathbf{k}_{\parallel})$ to the SOI due to the tilt $\delta\theta$ of the QW plane relative to the (111) orientation~\footnote{See Supplemental Material at [URL] for detailed information about the model for the electron spin lifetime in (111) QWs.}. The  lines in Fig.~\ref{spinlifetime}a show the  dependence of $\tau_z$ calculated  assuming $\tau^*_p=5$~ps and a ratio $r_{41}/\gamma=0.35$. These parameters, which determine $E_c$ and the peak lifetime, respectively,  were selected to match the experimental data for the SQW (the measured $E_c$ for the MQW may be affected by an inhomogeneous field distribution within the layer stack~\footnote{See Supplemental Material at [URL] for detailed information about the effect of inhomogeneous field distributions on the measured spin lifetime}).
Note that the model predicts reasonably well  the decrease of  $E_c$ with increasing temperature  (cf.~Eq.~\ref{EqEc}) as well as the temperature dependence of the  lifetime at $E_c$. Finally, by using  $\gamma=17\pm2$~eV{\AA}$^3$ determined in Ref.~\onlinecite{PVS158}, we obtain  $r_{41} = 6\pm1$~e{\AA}$^2$. The latter is   slightly  above $r_{41}= 5.2$~e{\AA}$^2$ calculated in Ref.~\onlinecite{Winkler03a} and $r_{41}=4$~e{\AA}$^2$ measured in Ref.~\onlinecite{Eldridge_PRB77_125344_08}.

The reverse field is also expected to increase the lifetime $\tau_R$ of spins precessing around an in-plane magnetic field B$||y$. This behavior is demonstrated in Fig.~\ref{spin2}, which  compares  $\rho_s(t)$ traces  measured in the SQW sample under $B=0$ (dashed lines) and $B=120$~mT (solid lines). Both $\tau_z$ and $\tau_R$ increase under moderate reverse biases. The initial spin polarization reduces significantly under a magnetic field. This behavior arises from the fact that the temporal width of the laser pulse (of approx. 600~ps) is comparable to the spin precession period (of approx. 1.5~ns). Finally, the value of the g-factor obtained from the fitting of the precessing spins at zero bias is $|g|\approx0.42$~\footnote{See Supplemental Material at [URL] for bias dependence of the g-factor}, in good agreement with that expected for electrons confined in a thick GaAs QW.

Figure~\ref{spinlifetime}b displays the field dependence of the lifetime $\tau_R$ of precessing spins in the SQW. In the range of investigated magnetic fields, we have checked that  $\tau_R$ for this sample is independent of the magnetic field amplitude, thus ensuring that it is not limited by inhomogeneities in the g-factor (the same is not expected for the MQW sample~\footnote{See Supplemental Material at [URL] for detailed information about the effect of inhomogeneous field distributions on the measured spin lifetime}).  The electric field dependence of $\tau_R$ in Fig.~\ref{spinlifetime}(b) can be divided into two regions. For applied fields away from $E_c$, the spin lifetime increases under a magnetic field.  This behavior can be understood from Eq.~1 by taking into account that at low temperatures $\Omega_{\mathrm{SO},z}(\mathbf{k}_{\parallel}) << \Omega_{\mathrm{SO},x}(\mathbf{k}_{\parallel}) \approx \Omega_{\mathrm{SO},y}(\mathbf{k}_{\parallel})$. Under this condition, $\tau_x\simeq\tau_y =2\tau_z$ and $\tau_R=\left[(1/2)\left(\tau^{-1}_x+\tau^{-1}_z \right)\right]^{-1} \approx 4\tau_z/3$.

Close to $E_c$, in contrast,  the  z-component of $\mathbf{\Omega}_\mathrm{BIA}(\mathbf{k}_{\parallel})$ (cf.~Eq.~1), which is not compensated by the SIA term, leads to the $\tau_R<\tau_z$.
The dependence of the spin lifetime on bias, temperature, and magnetic field is well reproduced by the dashed lines, which display the lifetime of precessing spins calculated using the model described in connection with Eq.~\ref{Eqtaui}.
Note, however, that the temperature dependence of the peak lifetimes in Fig.~ \ref{spinlifetime}(b) is weaker than that expected from the cubic k-terms of Eq.~1. Also, the model cannot reproduce the width of the $\tau_z$ and $\tau_R$ peaks  as a function of the electric field nor account for the  different lifetimes for the SQW and MQW in Fig.~\ref{spinlifetime}(a). These discrepancies indicate the existence of additional scattering channels for precessing spins.

In conclusion, we have experimentally demonstrated the electric control of the spin lifetime in GaAs(111) QW embedded in a n-i-p structure. In particular, the  Rashba SOI field induced by  the applied bias  can  compensate the Dresselhaus contribution, leading to spin lifetimes exceeding 100~ns. The experimental findings are well reproduced by a model for the spin relaxation mechanism, which yields the  Rashba and Dresselhaus spin-splitting coefficients.  The results presented here thus establish GaAs(111) QWs as excellent structures for the electrical storage and manipulation of spins.

We thank Dr. Flissikowski for a discussion as well as M. H\"oricke and S. Rauerdink for MBE growth and sample processing. We gratefully acknowledge financial support from the German DFG (priority program  SSP1285).


%

\end{document}